\documentclass{ethpaper}
\usepackage{graphicx}
\usepackage{here}
\usepackage{amssymb}
\usepackage{amsmath}
\usepackage{subfigure}
\usepackage{lineno}

\begin{document}
\begin{titlepage}
\ethnote{}

\title{Performance of a Tungsten-Cerium Fluoride Sampling Calorimeter in High-Energy Electron Beam Tests}

\begin{Authlist}
R. Becker, L. Bianchini, G. Dissertori, L. Djambazov, M. Doneg\`a, W. Lustermann, \\A. C. Marini, F. Nessi-Tedaldi, F. Pandolfi$^{\dagger}$, M. Peruzzi, M. Sch\"onenberger
\Instfoot{eth}{Institut for Particle Physics, ETH Zurich, Switzerland}

F. Cavallari, I. Dafinei, M. Diemoz, C. Jorda Lope, P. Meridiani, M. Nuccetelli, R. Paramatti, F. Pellegrino
\Instfoot{roma_infn}{INFN -- Sezione di Roma, Italy}

F. Micheli, G. Organtini, S. Rahatlou, L. Soffi
\Instfoot{roma}{Dipartimento di Fisica ``Sapienza'', Universit\`a di Roma e INFN -- Sezione di Roma, Italy}

L. Brianza, P. Govoni, A. Martelli, T. Tabarelli de Fatis
\Instfoot{milano}{Dipartimento di Fisica, Universit\`a di Milano Bicocca and INFN  -- Sezione di Milano-Bicocca, Italy}

V. Monti, N. Pastrone, P. P. Trapani
\Instfoot{torino}{INFN -- Sezione di Torino, Italy}

V. Candelise, G. Della Ricca
\Instfoot{trieste}{Universit\`a degli Studi di Trieste e INFN -- Sezione di Trieste, Italy}

\vspace{1cm}

$^{\dagger}$ Corresponding autor\\Email: francesco.pandolfi@cern.ch
\end{Authlist}

\maketitle

\vspace{0.5cm}
\begin{abstract}
A prototype for a sampling calorimeter made out of cerium fluoride crystals interleaved with tungsten plates,
and read out by wavelength-shifting fibres, has been exposed to beams of electrons with energies between
20 and 150~GeV, produced by the CERN Super Proton Synchrotron accelerator complex. 
The performance of the prototype is  presented and compared to that of a {\sc Geant4} simulation 
of the apparatus. Particular emphasis is given to the response uniformity across the channel front face, and to the prototype's energy resolution.
\end{abstract}

\vspace{1cm}
{\em Submitted to Nuclear Instruments and Methods A}

\end{titlepage}

\section{Introduction}

Future hadron colliders are expected to have high instantaneous luminosity~($> 10^{34}$~cm$^{-2}$~s$^{-1}$) and will operate for long data-taking periods, delivering to experiments large quantities of data~($> 1$~ab$^{-1}$).
This high-radiation environment sets stringent requirements on the radiation hardness of all detector parts,
and on the components placed in the forward region in particular.

A sampling calorimeter made of cerium fluoride~(CeF$_3$) scintillating crystals, interleaved with tungsten
plates, constitutes a viable option for forward electromagnetic calorimetry. Cerium fluoride spontaneously recovers, at room temperature, from hadronic damage~\cite{r-NIMCEF3}, and its stoichiometry can be tuned
in order to make it extremely resistant to ionizing radiation~\cite{r-EACEF3}. These features, together
with the fact that its scintillation light has a spectrum which is suitable for wavelength 
shifting~(the scintillation peak is around 340~nm) and is fast enough for the high-repetition frequencies of modern colliders~(it has a decay
time of 30~ns), make cerium fluoride a very attractive active material.

Stemming from the design of a previously tested prototype~\cite{btf}, a new design has been built, 
consisting of a single tower of 15 layers of alternating CeF$_3$ crystals and tungsten absorber plates. 
This article describes the results of irradiating this prototype with electrons of energies between 20 and 150~GeV, produced by the CERN Super Proton Synchrotron~(SPS) accelerator complex.

\begin{figure}[b]
  \centering
  \includegraphics[width=0.59\textwidth]{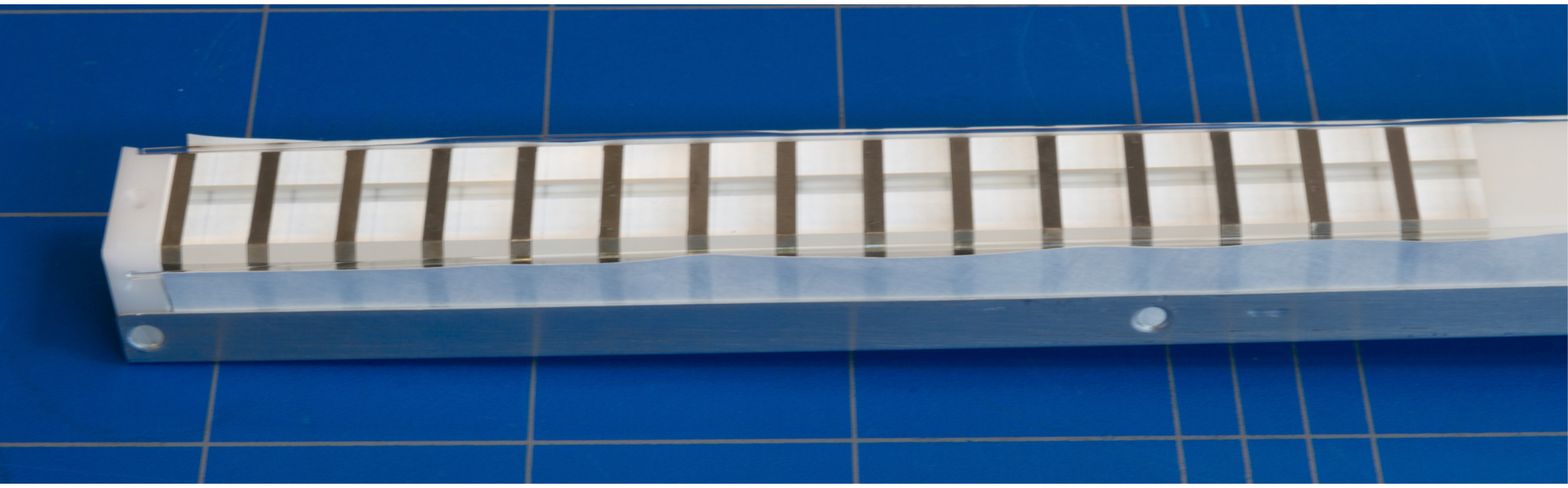}
    \includegraphics[width=0.4\textwidth]{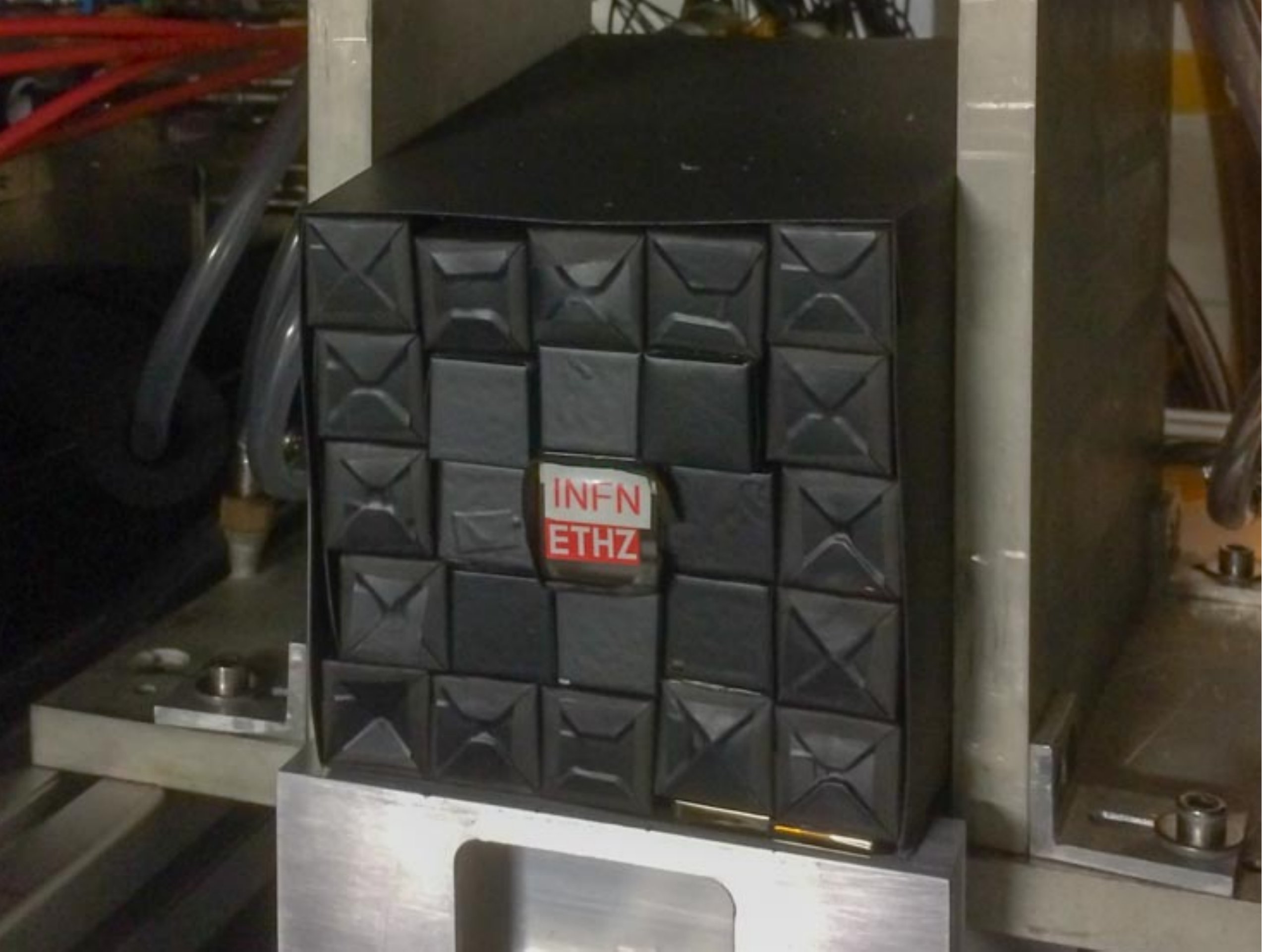}
  \caption{Left: The sampling calorimeter tower during assembly.
                Right: The front of the assembled $5\times5$ matrix.
  \label{fig:prototype}}
\end{figure}

\section{The W-CeF$_3$ Prototype}

A picture of the prototype during assembly can be seen in Figure~\ref{fig:prototype}~(left). It is made of 15 layers of 10~mm-thick CeF$_3$ crystals, interleaved with 3.1~mm-thick tungsten absorber plates, for a total length of $25\,X_0$.  The Barium-doped CeF$_3$ crystals have been produced by Tokuyama~\cite{tokuyama}, and their faces have been polished for total internal reflection.  The module transverse dimension of $24 \times 24$~mm$^2$ matches the effective Moli\`ere radius of the sampling compound, which is found to be 23.1~mm from a {\sc Geant4}~\cite{geant4} simulation.  A higher granularity can be achieved by increasing the absorber fraction, at the cost of lower sampling fraction.

Four 3~mm-wide chamfers are cut on the short edges, allowing wavelength-shifting~(WLS) fibres to run alongside the tower, transmitting the scintillation light to the back of the detector. The chamfers have been lapped to a roughness of $R_a = 0.4\pm0.1$~$\mu$m, so as to allow the light to escape the crystals and be collected by the WLS fibres. A 0.2~mm layer of highly-diffusive DuPont\texttrademark Tyvek\textsuperscript{\textregistered}~\cite{tyvek} foil is inserted between each crystal and its neighboring absorber plates, in order to maximise the light collection of the fibres. 

In this beam test four 3HF fibres produced by Kuraray~\cite{kuraray} have been employed. The fibres have a peak excitation of 340~nm, compatible with the emission spectrum of the CeF$_3$, and an emission peak at 540~nm. The fibres have a single 0.02~mm layer of cladding made of Polymethyl Methacrylate~(PMMA) surrounding the core of Polystyrene. One end is coupled with optical couplant to a Hamamatsu~\cite{hamamatsu} R1450 photomultiplier tube (PMT) at the rear end; the other end is aluminized. The quantum efficiency of the PMTs for the light emitted by the fibres has been measured to be about 7\%.
The radiation hardness of these fibres is not sufficient for future applications at the HL-LHC, therefore R\&D targeted at developing radiation-hard fibres is currently ongoing.

For lateral shower containment, the W-CeF$_3$ prototype has been surrounded by 24 BGO crystals, to complete a $5 \times 5$ matrix. The BGO crystals have been taken from the disassembly of the electromagnetic calorimeter of the L3 experiment~\cite{l3} operating at LEP, and are 24~cm long and have a front face of $22 \times 22$~mm$^2$ and a rear face of $30\times 30$~mm$^2$. The light from each BGO crystal is read out with a Hamamatsu R1450 PMT. A photograph of the front of the assembled $5\times 5$ matrix is found in Figure~\ref{fig:prototype}~(right). The matrix was  put in a light-tight box, and kept at a temperature of $18.0 \pm 0.1^{\circ}$~C. The box was placed on a moving table, which allowed to center the beam anywhere on the matrix, with a precision of 0.1~mm.

The fibre signals were then read out by a CAEN V1742 digitizer \cite{caen}, while the BGOs were read out by a CAEN V792 ADC.
The digitizer gives access to the full pulse-shape of the signal, which is stored in 1024 samples, at a frequency of 5~GHz.
The width of the readout window allowed a baseline subtraction on a per-event basis.
A low-pass filter, which was found to reduce high-frequency noise present in the signal and improve the stability of the measurement, was applied to the output of the digitizer. 
The channel response is defined as the full charge-integrated signal, after the baseline subtraction.

\section{Experimental Setup and Electron Selection}

The high-energy electrons from the H4 beam at the CERN SPS North Area are tracked before hitting the front face of the W-CeF$_3$ prototype: a wire-chamber~(placed at 12~m from the channel), two scintillating-fibre hodoscopes~(at 6 and 3~m, respectively), and a small hodoscope, just in front of tower front face. 

The wire chamber is made of two planes of 55 cathode wires and 28 anode wires, organized in a grid with an active area of $80\times 80$~mm$^2$. Each plane provides, respectively, a measurement of the electron position in the $x$ and $y$ directions with a nominal resolution better than 200~$\mu$m~\cite{wc}.

Each of the two scintillating-fibre hodoscopes is composed of two layers of 64 plastic fibres of 0.5 mm diameter, oriented in the $x$ and $y$ directions, respectively. The signals from the fibres are clustered grouping together adjacent fibres, up to a maximum of 4 fibres per cluster. The cluster position is defined as the average position of its fibres and it is used to estimate the trajectory of the particle before impacting the calorimeter. 


Finally, a small $2\times 2$~mm$^2$ hodoscope is placed in front of the calorimeter front face, and aligned with the center of the CeF$_3$ module. This hodoscope is made of two pairs of 1-mm fibres, again oriented in the $x-y$ directions. Being fixed to the calorimeter box, this small hodoscope is used to align the center of the tower front face with respect to the tracking detectors: events are selected in which the particle produces a signal in one of the two fibres of the small hodoscope, and the average recorded position is measured in each of the other three tracking devices. The average position measured by each tracking device is then used as an alignment offset, and is subtracted from subsequent measurements. This is done separately for each beam energy. After the alignment procedure the hodoscopes and the wire chamber are aligned within 0.5~mm.

Four plastic scintillators of varying size, the smallest of which has a transverse dimension of $1\times 1$~cm$^2$, are placed on the beamline. The data acquisition is triggered by the coincidence of all four scintillators.

Beam electrons are found to most likely produce 2-fibre clusters. Electrons hitting upstream material will produce particle showers before hitting the tower; these events are vetoed by selecting only events 
which registered hodoscope activity compatible with the passage of one electron.
It is also additionally required that electrons are parallel to the beam line
by requiring that the position reconstruction of the wire chamber and the far hodoscope are within 4~mm of each other, and that the positions measured in the two hodoscope planes lie within 1.5~mm.
The reconstructed impact position of the electron on the calorimeter is taken as the average of the positions measured by the two hodoscopes.

\begin{figure}[tb]
  \centering
  \includegraphics[width=0.49\textwidth]{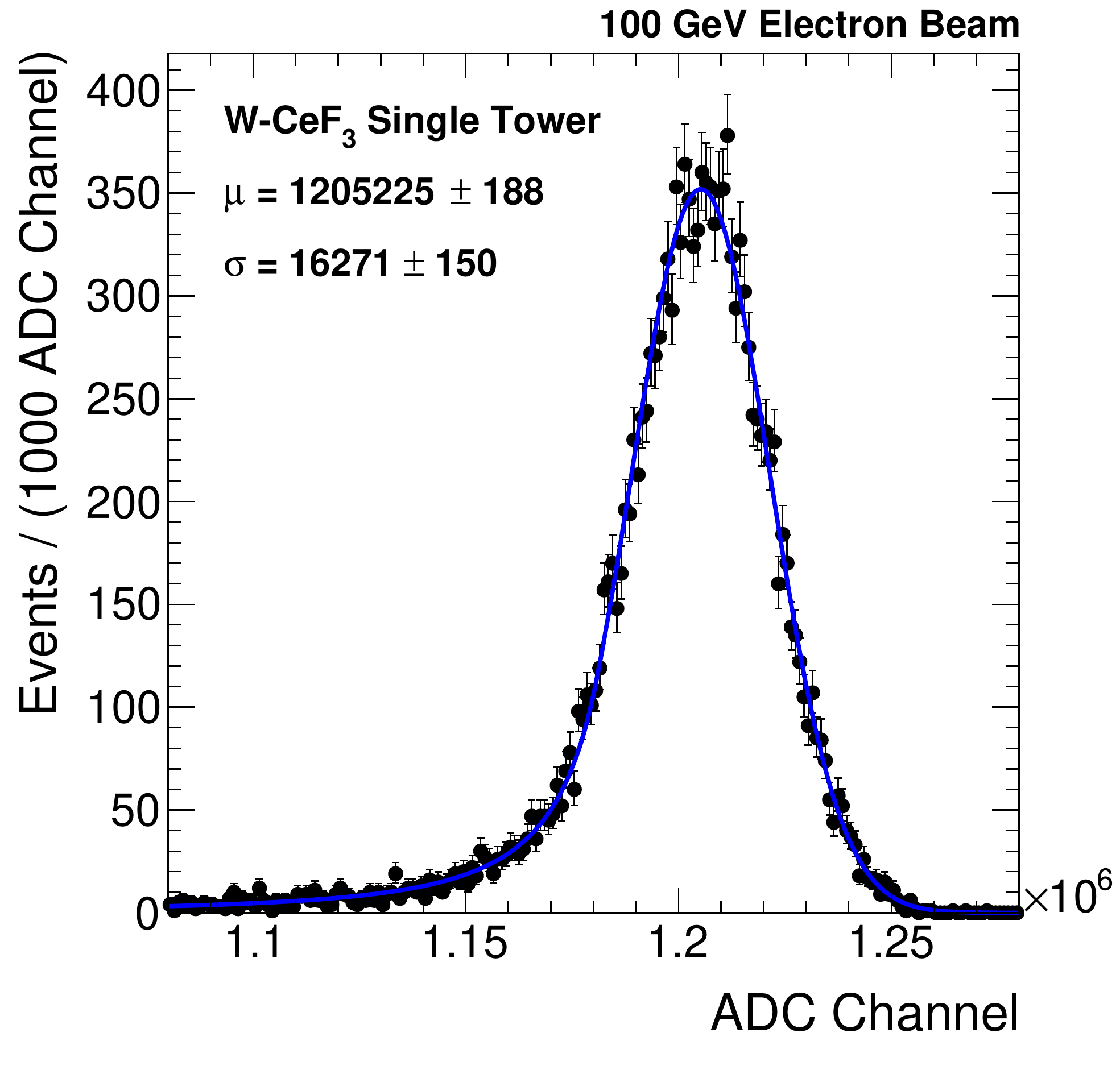}
  \caption{Example response distribution, for 100 GeV electrons. The result of a Crystal-Ball fit is superimposed with a blue line, and the mean~$\mu$ and gaussian width~$\sigma$ of the fitted function are shown.
  \label{fig:crystalball}}
\end{figure}


\section{Experimental Setup Simulation}

A full simulation of the experimental setup has been implemented within the {\sc Geant4} framework. The W-CeF$_3$ channel description has a realistic geometry of the CeF$_3$, tungsten, and the fibres. The BGO crystals are approximated by 22~mm~$\times$~22~mm~$\times$~24~cm parallelepipeds. The hodoscopes and scintillators along the beam line are approximated with blocks of PMMA with the corresponding sizes.

The simulated response of the channel is taken as the sum of the energy deposited in the active material, and each layer is weighted by the relative light yield difference of the 15 crystals, which has been measured in the laboratory with a radioactive source, and  found to have a spread of 9\%. The simulation automatically takes into account resolution effects due to stochastic shower fluctuations and lateral containment. An additional term to include photostatistics, determined by the {\em in situ} absolute photoelectron calibration from cosmic muons, is added in quadrature.

In addition to the experimental setup, an additional setup, consisting of a $5\times 5$ matrix of W-CeF$_3$ towers, has been simulated. This allows for a more realistic description of the lateral leakage and for an extrapolation to future calorimetric applications.

\begin{figure}[tb]
  \centering
  \includegraphics[width=0.49\textwidth]{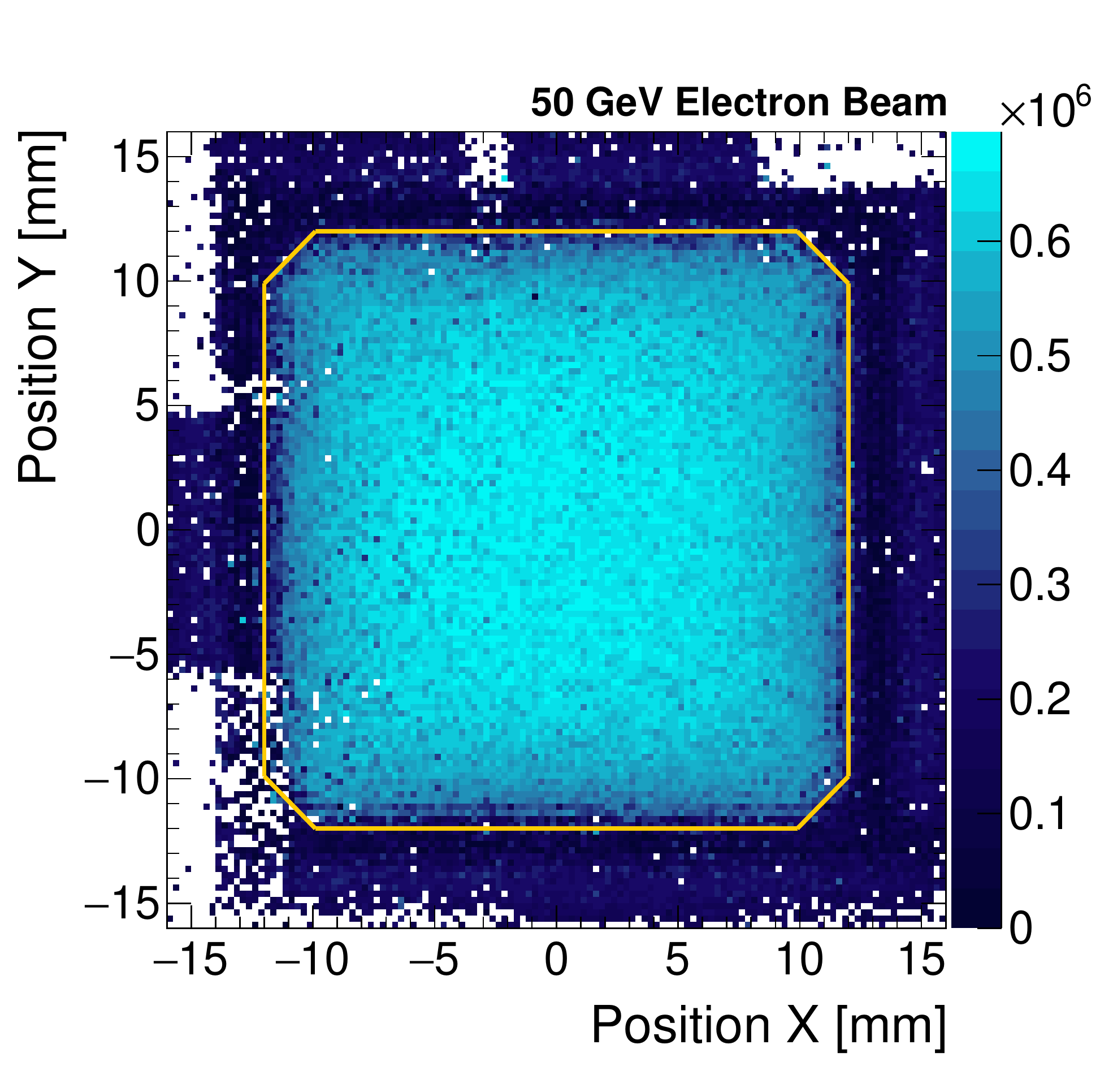}
  \includegraphics[width=0.49\textwidth]{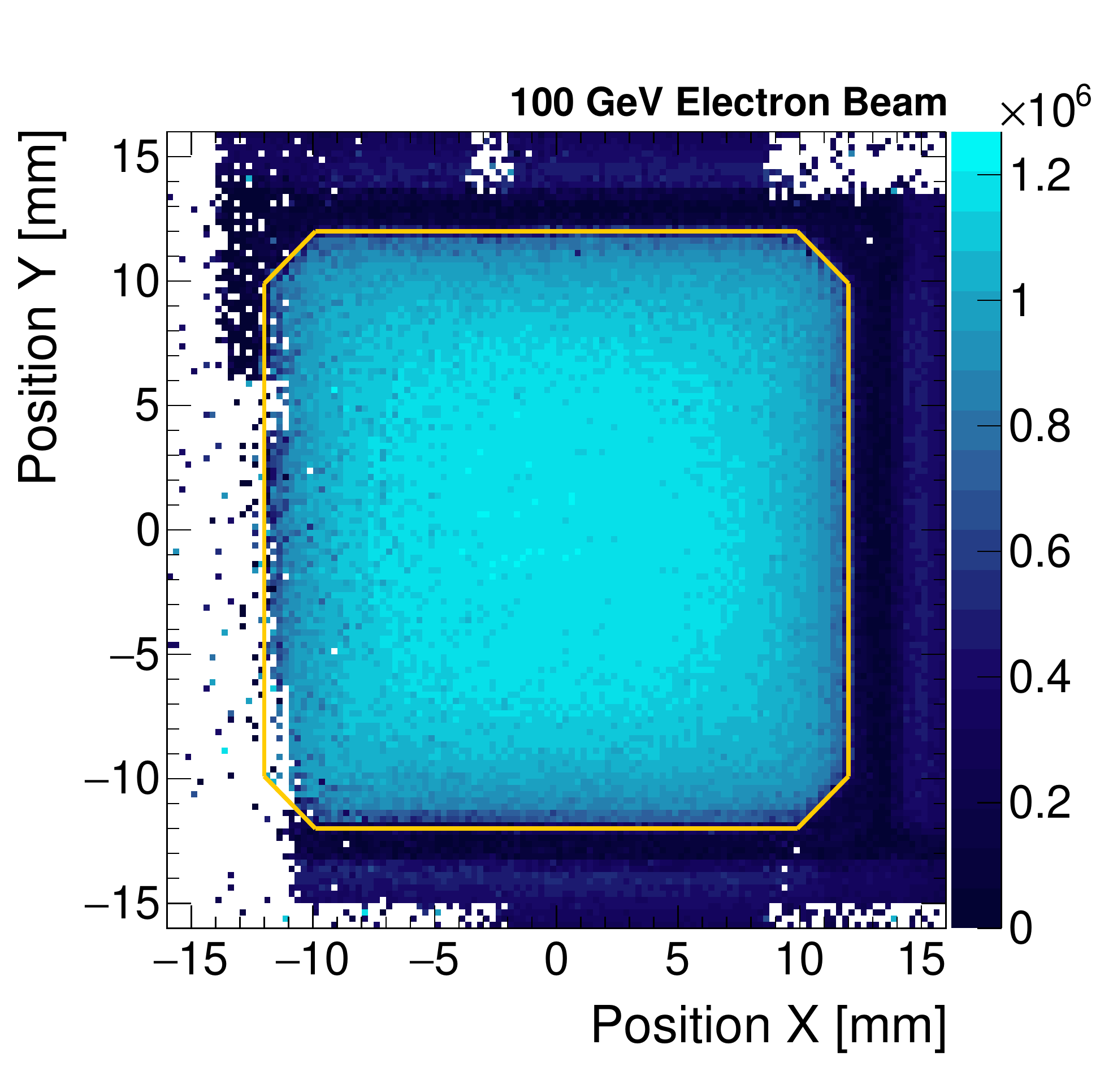}
  \caption{Response of the W-CeF$_3$ channel as a function of the electron impact point on the front face for 
50~GeV~(left) and 100~GeV~(right) electrons.
  \label{fig:resp_ff}}
\end{figure}

\begin{figure}[tb]
  \centering
  \includegraphics[width=0.49\textwidth]{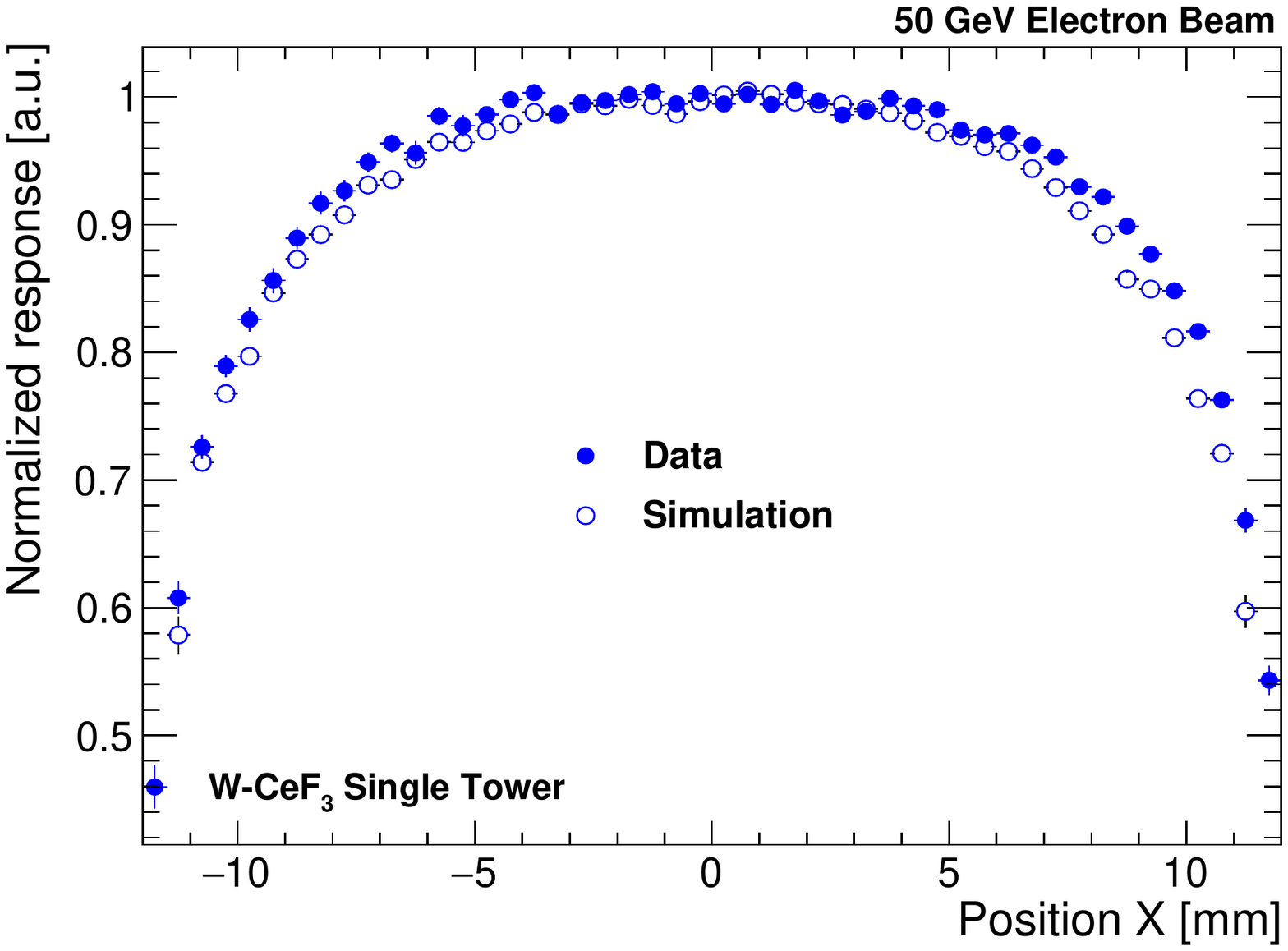}
  \includegraphics[width=0.49\textwidth]{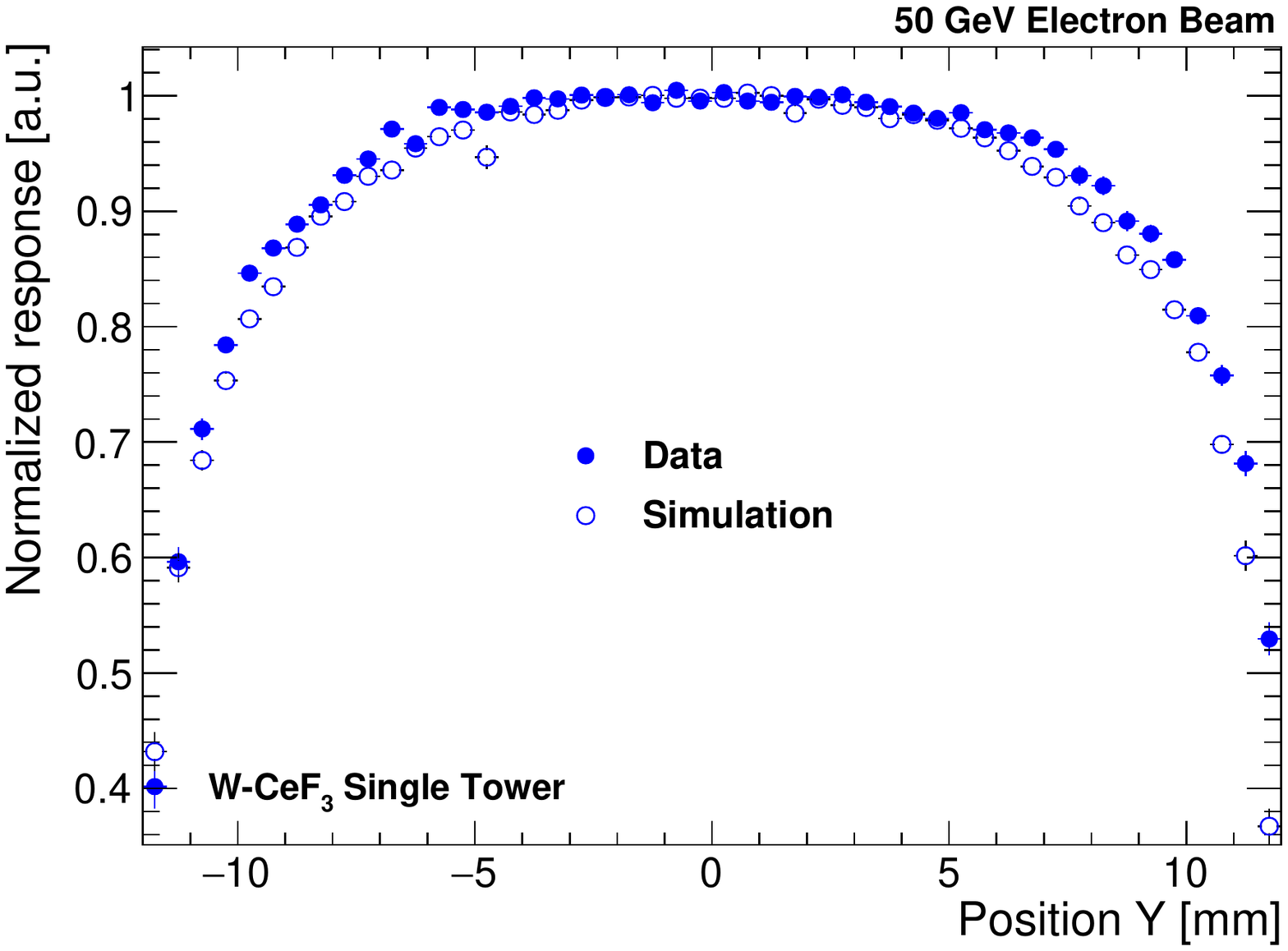}
  \caption{Response profile as a function of the beam impact position, projected along the $x$~(left) and $y$~(right) directions, for data~(solid markers) and the simulation~(open markers).
  \label{fig:resp_vs_mc}}
\end{figure}

\section{Results}
\label{sec:results}

The response of the four read-out channels of the prototype is intercalibrated in order to take into account  differences in the gains and quantum efficiencies of the PMTs. This is done by selecting events where electrons impinge on a narrow $\pm3$~mm strip passing through the center of the front face, alternatively in the $x$ and $y$ directions, and studying the variation of the response of each single read-out 
channel as a function of the electron impact position along that strip. The inverse of the response at the center of the strip is taken as intercalibration factor, and the two factors obtained in the $x$ and $y$ directions are averaged. The resulting intercalibration factors are applied as a multiplicative, relative, correction. The four calibration factors are found to be within 10\% of unity.

The response of the W-CeF$_3$ channel is defined through a fit to the sum of the inter-calibrated signals recorded by the four fibres. The adopted fit function is a Crystal Ball~\cite{crystalball1, crystalball2}, to model the low-energy response tail. An example fit, for 100~GeV electrons, can be seen in Figure~\ref{fig:crystalball}. 

The resolution of the tracking devices on the beam allows for a precise study of the prototype's response as a function of the electron impact point on the front face. This is shown in Figure~\ref{fig:resp_ff} for 
50~GeV~(left) and 100~GeV~(right) electrons. 
The nominal position of the channel's front face is overlayed as a yellow frame, and the agreement with the observed response profile is proof of a successful alignment. Furthermore, the response is found to be quite uniform across the channel's front face. 

Figure~\ref{fig:resp_vs_mc} shows instead the response profile as a function of the beam impact position, projected along the $x$~(left) and $y$~(right) directions, respectively for data~(solid markers) and the simulation~(open markers). 
The simulation does not include ray tracing and light collection effect. Nevertheless the difference between the data and simulation is at most 5\%, so we can conclude that the effect of light collection in the four fibers is smaller than 5\% in this configuration.  In order to minimize the effect of lateral containment fluctuations, for the resolution studies that follow, events impacting the central $6\times 6$~mm$^2$ area of the channel front face have been selected.



The resolution of the response distribution $\sigma_R$ is defined as the ratio between the mean and the width of the gaussian core of the Crystal Ball fit. To extract the energy resolution $\sigma_E$ of the W-CeF$_3$ tower, it is necessary to disentagle (i) PMT saturation effects~(present at high voltage settings) and (ii) PMT noise~(present at low voltage settings). Because of the dependence of both effects on the high voltage applied to the PMTs, in situ voltage scans have been taken at different beam energies.

\begin{figure}[tb]
  \centering
  \includegraphics[width=0.49\textwidth]{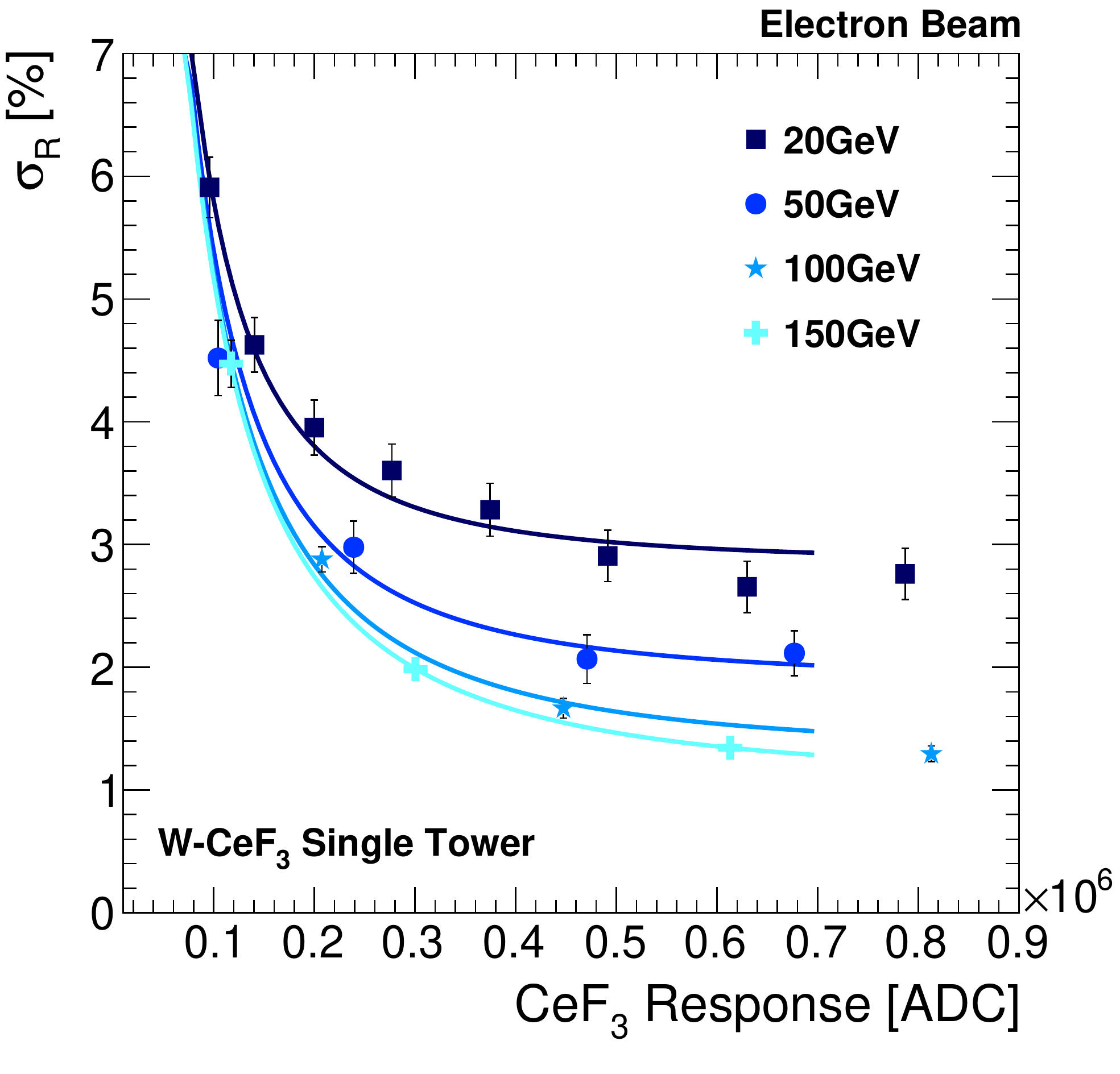}
  \caption{Response resolution~$\sigma_R$ as a function of the response, for beam energies of 20~GeV~(squares), 50~GeV~(circles), 100~GeV~(stars) and 150~GeV~(crosses). The results of the fits are superimposed.
  \label{fig:sigma_vs_R}}
\end{figure}

The result of these scans is shown in Figure~\ref{fig:sigma_vs_R}, where $\sigma_R$ is studied as a function of the mean response $R$, for beam energies of 20~GeV~(squares), 50~GeV~(circles), 100~GeV~(stars) and 150~GeV~(crosses). The points are fitted with the function:
$$
\sigma_R(R) = \frac{N}{R} + \sigma_E
$$
where the first term describes PMT noise and is governed by the parameter~$N$ which is fitted simultaneously to all energy response scans, and the second term is the instrinsic energy resolution, at a given energy. The results of the fits are superimposed in Figure~\ref{fig:sigma_vs_R}.  The uncertainties of the points shown in Figure~\ref{fig:sigma_vs_R} include a systematic uncertainty related to the choice of the functional form chosen to fit the response distributions, which has been evaluated separately for each energy, and reaches a maximum value of 20\% at the lowest energy point. 

In order to limit the effect of PMT saturation, which manifests itself at high response, the fitting range is limited to $0.7\times10^6$~ADC channels. The fit is repeated twice, with an upper limit of $0.5\times 10^6$ and $0.9\times 10^6$~ADC channels, respectively, and the maximal variation in $\sigma_E$ is taken as a systematic uncertainty on the choice of the fitting range. It is found to be less than~10\% for all energies.

\begin{figure}[tb]
  \centering
  \includegraphics[width=0.6\textwidth]{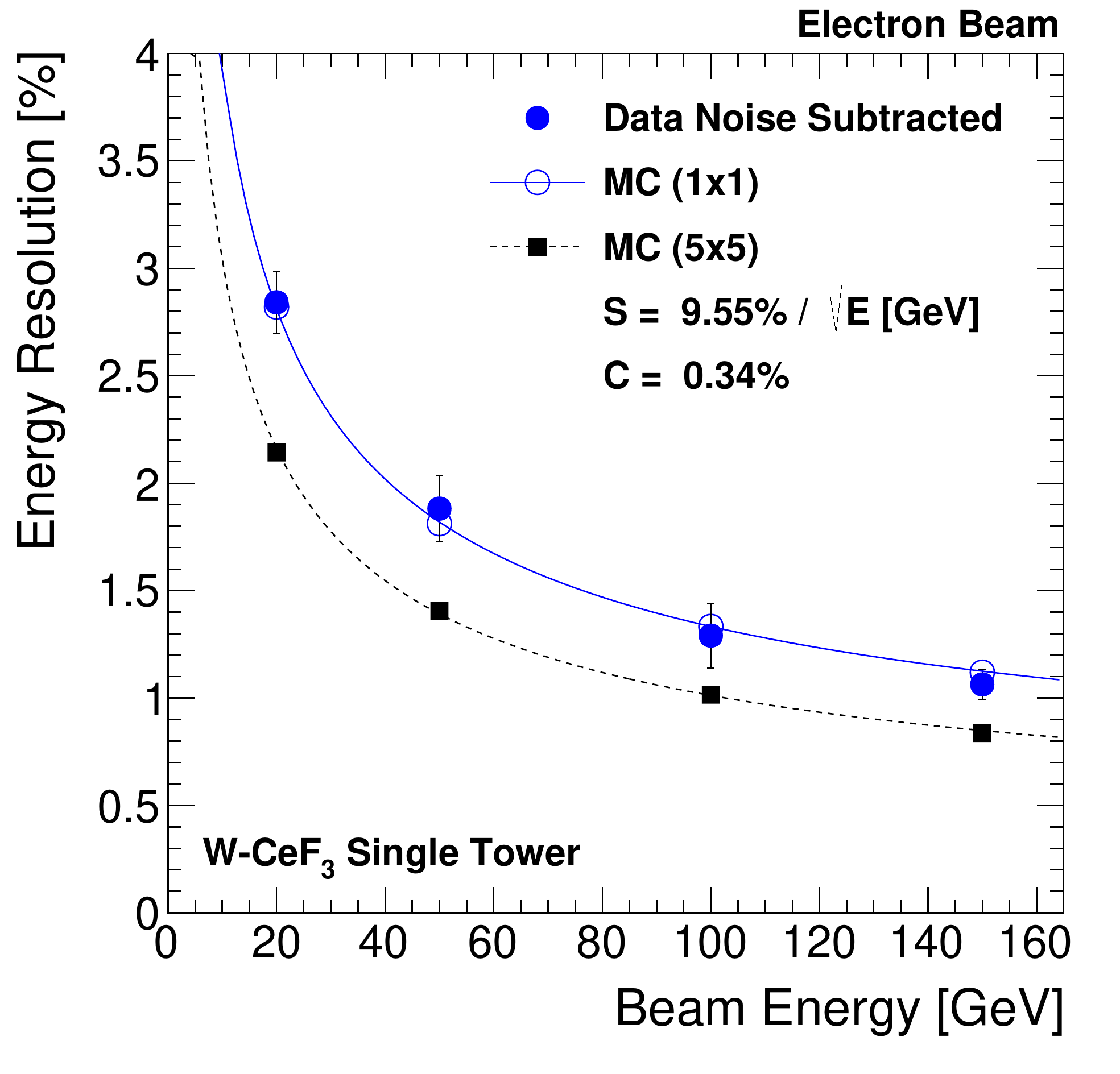}
  \caption{Energy resolution of the W-CeF$_3$ single tower as a function of the electron beam energy~(solid blue markers), compared to the simulation~(hollow markers). Resolution of the simulated~$5\times 5$ W-CeF$_3$ matrix is shown with square black markers.
  \label{fig:moneyplot}}
\end{figure}

The resulting energy resolutions~$\sigma_E$ of the prototype are shown with solid blue markers in Figure~\ref{fig:moneyplot}, as a function of the electron beam energy. 
The data are compared to the results obtained with the simulation~(hollow markers) and the two are found to be in agreement. This allows for an extrapolation to the simulated $5\times 5$ simulated matrix, whose results are shown with black square markers, and are fitted with the function:
$$
\sigma_E(E) = \frac{S}{\sqrt{E\,[\mathrm{GeV}]}} \oplus C
$$
where $S$ is the stochastic term and $C$ is the constant term. We obtain $S = 9.54 \pm 0.13\%$ and $C = 0.33 \pm 0.04\%$.

\section{Conclusions}

 A prototype for a sampling calorimeter made out of cerium fluoride crystals interleaved with tungsten plates,
and read out by wavelength-shifting fibres, has been exposed to beams of electrons with energies between
20 and 150~GeV, produced by the CERN Super Proton Synchrotron accelerator complex. 
The tower response is found to be uniform across the channel front face. 
We find the measured energy resolution to be in good agreement with a dedicated {\sc Geant4} simulation. According to the same simulation, a $5\times 5$ tower matrix would achieve a stochastic term of $(9.54 \pm 0.13)\%/ \sqrt{E\,[\mathrm{GeV}]}$. This constitutes a proof of concept of a technology with potential use in future hadron-collider applications.

\section*{Acknowledgments}
The excellent support of the ETH Zurich technical staff is acknowledged, in particular M. Dr\"oge, C. Haller and U. Horisberger, as is the one of CERN, by R. Dumps, T. Schneider and M. Van Stenis. We thank Alexander Singovski and Adrian Fabic for their precious assistance at the SPS North Area. We are grateful to A. Vedda, N. Chiodini and M. Fasoli (U. Milano Bicocca), and to Kuraray (Japan) for kindly providing us with samples for this test. We also acknowledge the effort made by Hamamatsu (Japan) and Tokuyama (Japan), to provide us with the needed parts in time for meeting a stringent construction schedule. This work was performed with the support of the Swiss National Science Foundation. 
\\





\begin{thebibliography}{1}

\bibitem{r-NIMCEF3} G.~Dissertori, P.~Lecomte, D.~Luckey, F.~Nessi-Tedaldi, F.~Pauss, 
  Th.~Otto,  S.~Roesler, Ch.~Urscheler,
Nucl.~Instr.~and Meth.~Phys.~Res.~A 622 (2010) 41-48.

\bibitem{r-EACEF3} E.~Auffray, S.~Baccaro, T.~Beckers, Y.~Benhammou, A.~N.~Belsky, B.~Borgia, D.~Boutet, R.~Chipaux, et al., Nucl.~Instrum.~and Meth.~Phys.~Res.~A 383 (1996) 367-390.

\bibitem{btf} R.~Becker, et al., {\em Beam test results for a tungsten-cerium fluoride sampling calorimeter with wavelength-shifting fiber readout}, submitted to JINST

\bibitem{geant4} S.~Agostinelli et~al., Nucl. Instrum. and Meth. Phys. Res. A 506 (2003) 250-303.

\bibitem{tokuyama} Tokuyama Corporation, 1-1, Mikage-cho, Shunan city, Yamaguchi 745-8648 (Japan).

\bibitem{tyvek} Tyvek\textsuperscript{\textregistered} is a product of DuPont Inc., Wilmington, DE 19801-5003 (USA).

\bibitem{kuraray} Kuraray, Japan, 3HF scintillation fibers technical specifications at http://kuraraypsf.jp/psf/sf.html, visited on May 15th, 2014.

\bibitem{hamamatsu} Hamamatsu Photonics K.K., Hamamatsu City (Japan).

\bibitem{l3} L3 Collaboration, Nucl. Inst. Meth. A 289 (1990) 35


\bibitem{caen} CAEN S.p.A., Via Vetraia 11, 55049 - Viareggio (LU) - Italy.

\bibitem{wc} J.Spanggaard. Delay Wire Chambers - A Users Guide (CERN, Geneva, 1998).
URL {\tt http://cds.cern.ch/record/702443/files/sl-note-98-023.pd}.

\bibitem{crystalball1}  M. Oreglia, {\em A study of the reactions $\psi^{\prime} \rightarrow \gamma\gamma\psi$}, PhD thesis, Stanford University (1980) SLAC Report SLAC-R-236

\bibitem{crystalball2} G.I. Kirkbride et al., {\em IEEE Trans. Nucl. Part. Sci.}, Vol. NS-26 No.1~(1537), 1979.



\end{thebibliography}
%

\end{document}